\documentclass[prb,twocolumn,showpacs,amssymb]{revtex4}

\usepackage{graphicx}

\begin{document}

\author{Victor Kagalovsky and Demitry Nemirovsky}
\affiliation{Sami Shamoon College of Engineering,
Beer-Sheva, 84100 Israel}

\title{Universal critical exponent in class D superconductors}

\begin{abstract}
We study a physical system consisting
of non-interacting quasiparticles in disordered
superconductors that have neither time-reversal nor spin-rotation
invariance. This system belongs to class D within the 
recent classification scheme of random matrix ensembles (RME) and its phase diagram contains three different phases: metallic and 
two distinct localized phases with different quantized thermal Hall conductances.
We find that critical exponents describing different transitions (insulator-to-insulator and insulator-to-metal) 
are identical within the error of numerical calculations and also find that 
critical disorder of the insulator-to-metal transition is energy independent. 
\end{abstract}


\maketitle

The properties of quasiparticles in disordered superconductors have been intensively investigated. Those systems are representatives of new 
symmetry classes different from the familiar three classes in normal disordered conductors and in the Wigner-Dyson random matrix ensembles. A list 
of additional random matrix ensembles, determined by these new symmetry classes has been established about 10 years ago \cite{Zirn}. 
These additional random matrix ensembles describe 
zero-dimensional problems, and are appropriate to model a small grain of a 
superconductor in the ergodic highly conducting limit. 
In the corresponding higher-dimensional systems of the same symmetry classes there can be transitions between metallic, 
localized, or quantized Hall phases for 
quasiparticles \cite{Sent1, WEPRL, SENT, We}.

The symmetry class D (which we address in this paper) may be 
realized in superconductors with broken time-reversal invariance, and either 
broken spin-rotation invariance (as in d-wave superconductors with spin-orbit 
scattering) or spinless or spin-polarized fermions (as in certain p-wave
states). The associated changes in quasiparticle dynamics must be probed by 
energy transport, since neither charge density nor spin are conserved. 
In this paper we present extensive numerical results on a model first introduced by Cho and Fisher (CF) \cite{Cho} (see a detailed description below), which  
has a particularly rich phase diagram in two dimensions \cite{Senthil,We}. 
We use advanced numerical calculations, proposed in Ref. \onlinecite{We} and described in detail 
in \cite{Merz,Var} to overcome round-off errors in calculations of renormalized localization lengths.  
We also apply, for the first time,  an optimization algorithm to determine both critical exponent and critical disorder for 
the insulator-to-metal transition and critical exponent for
the insulator-to-insulator transition. Our results suggest the existence of the universal critical exponent $\nu=1.4\pm0.2$ for the thermodynamical localization length
$\xi$ for both types of transitions mentioned above, an energy independent value of the critical disorder $W_{cr}\approx 0.19\pm0.02$ and 
show collapse of all data on one curve 
with thermodynamic localization length $\xi\sim[\epsilon(W_{cr}-W)]^{-\nu}$.

The original network model \cite{CC} was proposed to describe transitions 
between plateaux in the quantum Hall effect (QHE). QHE is realized in a 
two-dimensional electron gas subjected to a strong perpendicular magnetic 
field and a random potential. When random potential varies smoothly 
(its correlation length is much larger than the magnetic length) a semiclasscial 
description becomes relevant: electrons move along the lines of constant 
potential. When two equipotential lines come close to each other (near a saddle 
point), tunneling is feasible.  In the network model, electrons move along 
unidirectional links forming closed loops in analogy with semiclassical 
motion on contours of constant potential. Scattering between links is allowed 
at nodes in order to map tunneling through saddle point potentials. 
Propagation along links yields a random phase $\phi$, thus links are 
presented by diagonal matrices with elements in the form $\exp (i\phi)$. 
Transfer matrix for one node relates a pair of incoming and outgoing amplitudes 
on the left to a corresponding pair on the right; it has the form
\begin{equation}
{\bf T}=\left( \begin{array}{cc}\cosh\theta & \sinh\theta  \\
 \sinh\theta & \cosh\theta
\end{array}
\right).
\label{first}
\end{equation}

The node parameter $\theta$ is related to the electron energy in 
the following way
\begin{equation}
\epsilon =-\frac{2}{\pi}\ln(\sinh\theta ),
\label{ninth}
\end{equation}
where $\epsilon$ is a relative distance between the electron energy and the 
barrier height. It is easy to see that the most "quantum" case (equal
probabilities to scatter to the left and to the right) is at $\epsilon =0$ 
($\theta =0.8814$), in fact, numerical calculations show that there is an 
extended state at that energy. 
 
Numerical simulations on the network model are performed on a system with fixed width $M$ and periodic boundary conditions 
in the transverse direction. 
Multiplying transfer matrices for $N$ slices and then diagonalizing the 
resulting total transfer matrix, it is possible to extract the smallest Lyapunov 
exponent $\lambda$ (the eigenvalues of the transfer matrix are $\exp(\lambda N)$). The 
localization length $\xi_M$ is proportional to $1/\lambda$. Repeating 
calculations for different system widths and different energies it is 
possible to show that the localization length $\xi_M$ satisfies a scaling
relation
\begin{equation}
\frac{\xi_M}{M} =f\left(\frac{M}{\xi (\epsilon )}\right) .
\label{second}
\end{equation}

When renormalized localization length $\xi_M/M$ becomes $M$-independent it strongly suggests that the system is in the critical state. Indeed, it means 
that when you double the system size $M$, the localization length $\xi_M$ also doubles, or, from the mathematical standpoint, the only way to make Eq. (3) $M$-independent 
and consistent is to put the thermodynamic localization legth $\xi$ equal to infinity (extended state).

For a D class symmetry 
a Bogoliubov - de Gennes Hamiltonian may be written in terms 
of a Hermitian matrix \cite{Zirn}. The corresponding time evolution 
operator is real, restricting the generalized phase factors to be 
O($N$) matrices for a model in which $N$-component fermions propagate on 
links, and to the values $\pm 1$ for $N=1$, the case that was studied. 
There are three models: random bond Ising model \cite{Choth,gruzb}, supporting two different localized phases, 
uncorrelated O(1) model \cite{boucq}, where phases on the links 
are independent random variables and all states are extended \cite{We}, and the model first introduced by Cho and 
Fisher (CF) \cite{Cho} where scattering phases with the value $\pi$ appear 
in correlated pairs. Each model has two parameters: 
the first one is  
a disorder concentration $W$, such that there is a probability $W$ ($1-W$) 
to have a phase $0$ ($\pi$) on a given link. The second parameter is an energy 
$\epsilon$ describing scattering at the nodes. For the CF model, the phase diagram (updated version of which is presented in Fig. 3) in the $\epsilon$-$W$ plane 
has three distinctive phases: metallic, and two insulating phases characterized 
by different Hall conductances. The sensitivity to the disorder is a distinctive 
feature of class D. 

The existence of region of extended states means that the smallest Lyapunov 
exponent at each particular energy is zero or extremely small. 
This point was discussed in detail in Ref. \onlinecite{We}. The advanced algorithm using the structure of the 
transfer matix imposed by current conservation and the symmetry of class D suggested there allowed to 
increase the accuracy of the data significantly. However, in the vicinity of the critical points the noise increases and in order to 
identify the values of critical disorder $W_{cr}$ at different energies one needs to focus only on large system sizes $M$ ,
therefore errors in the values of localization lengths $\xi_M$ can still be significant. In this paper we have decided to address 
this problem from the other direction. We use the values of $\xi_M$ we can trust (not very close to the critical point) and demand that 
all renormalized localization lengths satisfy scaling relations (allowing the program to find the optimal values of critical exponent and critical disorder).

In the CF model the disorder is introduced only at the nodes, allowing for the offdiagonal elements of Eq. (1) to be multiplied by $\pm 1$ (disorder probablity 
$W$ is the probability of that factor to be $-1$). 
We have started numerical calculation for fixed values of $W$ changing the system widths $M$ and the energies $\epsilon$. In those cases 
an obvious critical energy is $\epsilon =0$ (the most "quantum" case explained above) and we have been looking for the critical exponent $\nu$ which fits all the data onto one curve. We use a special 
optimization program which checks different critical exponents and chooses the optimal one. The typical data for the renormalized localization lengths 
for different system widths $M$ and energies $\epsilon$ for a fixed disorder $W=0.1$ is shown in Fig. 1a. For large $\epsilon$ the states 
are localizaed almost ideally (renormalized localization length 
is proportional to $1/M$). As one approaches the critical energy $\epsilon =0$, the values of all renormalized lengths increase significantly (tending to 
infinity). In Fig. 1b it is shown that all the data fit onto one curve when we scale the horizontal axis as $\epsilon M^{1/\nu}$ with the critical exponent of 
the thermodynamic localization length $\nu =1.4$. Indeed, substituting into the right-hand side 
of Eq. (3) $\xi\sim\epsilon^{-\nu}$ one can get the variable of the horizontal axis in Fig. 1b. We have performed such calculations for different fixed values of 
the disorder $W$ in the range $[0.05;0.15]$. The optimization program have found very close values of $\nu$ for all the data. 

\begin{figure}[t]
\begin{center}
\includegraphics[scale=0.85]{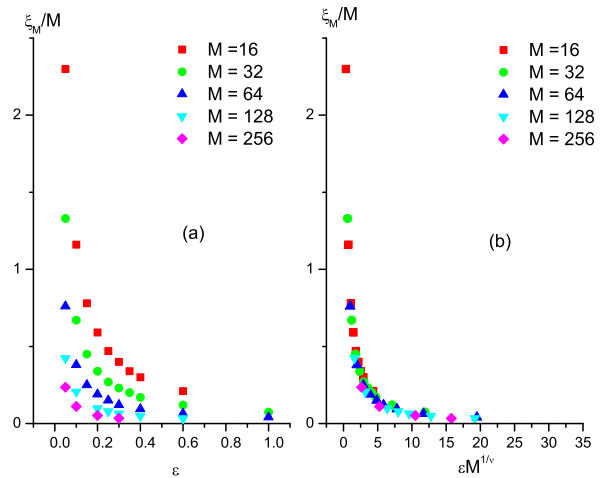}
\end{center}
\caption{(a) Renormalized localization lengths for various system widths $M$ and energies $\epsilon$ for a fixed disorder value
$W=0.1$.(b) Scaling of the same data with critical 
exponent $\nu=1.4$}
\end{figure}
  
Next, we address the most interesting problem - defintion of the critical disorder and corresponding critical exponent. We have fixed the value of energy
$\epsilon$ and vary $M$ and $W$. The renormalized localization lengths are shown in Fig. 2a for a fixed $\epsilon =0.1$. We then use the optimization program to find optimal values for both the critical disorder $W_{cr}$ and critical exponent. The results of the scaling are shown in Fig. 2b. Surprisingly,
the value of the critical exponent found in this way was identical (within the error of numerical calculations) as for a fixed disorder. 
The critical disorder found for a wide range of different fixed values of energy
$\epsilon$ 
($\epsilon =0.1, 0.2, ...1.$) 
has the same value $W_{cr}$ independent on $\epsilon$. 
This way of defintion of $W_{cr}$ 
by the otimization procedure looks more promising, especially taking into account that when one looks for $\xi_M/M$ to become $M$-independent 
in the standard procedure, the errors increase 
significantly and one needs to take $M$ as big as possible and to trust only to the largest values of $M$. We, therefore, suggest that our method 
leads to the phase digram shown in Fig. 3 with the vertical critical line separating insulator and metal (for energies $\epsilon$ not very close to zero). We, however, cannot determine the exact position of the multicritical point at $\epsilon =0$ (studied previously \cite{We,evers}), which is underlined 
by the dots in Fig. 3.

\begin{figure}[t]
\begin{center}
\includegraphics[scale=0.85]{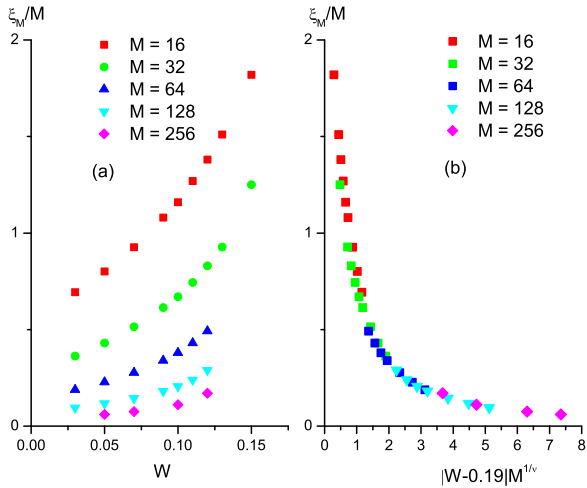}
\end{center}
\caption{(a) Renormalized localization lengths for various system widths $M$ and disorder $W$ for a fixed energy value
 $\epsilon=0.1$.(b) Scaling of the same data with critical 
exponent $\nu=1.4$ and critical disorder $W_{cr}=0.19$.}
\end{figure}

\begin{figure}[htb]
\begin{center}
\includegraphics[scale=0.5,bb= 14 12 268 242]{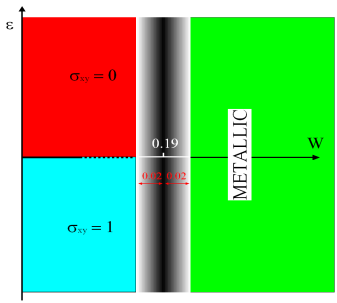}
\end{center}
\caption{Updated phase diagram for a CF model with metallic, insulating and quantized Hall phases.}
\end{figure}

The very fact that we have obtained almost identical values of the critical exponent $\nu =1.4\pm 0.2$ for both types of transition,  
insulator-to-insulator transition with 
$\xi\sim\epsilon^{-\nu}$ 
and insulator-to-metal transition with $\xi\sim (W_{cr}-W)^{-\nu}$ (which is rather 
different from anything obtained before, (e.g. two different critical exponents for a class C \cite{WEPRL, SENT} ) has led us to the last fit. We have suggested 
that the thermodynamic localization length should depend on the product of both distances of energy and disorder from the critical ones with the same 
critical exponent in the following way $\xi\sim[\epsilon(W_{cr}-W)]^{-\nu}$. Indeed, all the data for the renormalized localization lengths 
obtained for various $\epsilon$, $W$ and $M$ collapse on the same curve after suggested fit as shown in Fig. 4.

\begin{figure}[t]
\begin{center}
\includegraphics[scale=0.85]{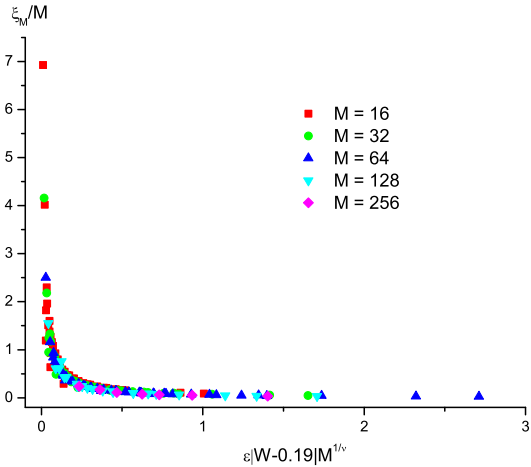}
\end{center}
\caption{Collapse of all of the data on a single curve with universal critical exponent $\nu=1.4$ and universal critical disoredr $W_{cr}=0.19$.}
\end{figure}

Detailed analytical explanations for the divergence of the localization length in various types of class D systems for 
a one-dimensional model ($M=2$) were presented earlier \cite{We,Var}. Here we wish to add a qualitative argument for a two-dimensional system. First, 
consider a system without disorder ($W=0$). Then, there are no phases on the links, tunneling alone defines a quaiparticle behaviour, and for any non-zero energy makes system an insulator. For any $\epsilon\neq 0$ at any node quasiparticle prefers to turn into the same direction circling closed trajectories. 
Futhermore, because the probability 
of a tunneling event is proportional to $\exp(-\pi\epsilon)$, it is trivial to derive that after $N$ 
tunneling events the probability becomes $\exp(-\pi\epsilon N)$, and, considering a number of tunneling events $N$ as a distance, one immediately obtains  
the localization length as $\xi\sim\epsilon ^{-1}$. We have found that those statements, mentioned above, are correct numerically as well. 
For $W=0$ and $\epsilon\neq 0$ all states are ideally localized ($\xi_M$ is $M$-independent) and fit onto one curve depending on 
$\xi/M$ with $\xi\sim\epsilon ^{-1}$. Now let us introduce disorder. It causes the appearence of phase $\pi$ on some links. Those phases 
can cause a constructive interference for different trajectories leading from one point of the network model to another, making 
therefore an extended state. For a $O(1)$ model phases on the 
different links are uncorrelated, and therefore even a small disorder $W$ introduces that constructive interference immediately. In the 
Cho-Fisher model that we consider in this paper, phases on both sides of the node are correlated (the same), therefore small disorder 
does not mean an immediate delocalization, but it is clear that the stronger the disorder is the larger is the opprotunity for the 
state to be extended. 

To summarize, we have studied critical exponents describing the divergence of thermodynamic localization lengths for two both types of transition 
(insulator-to-insulator and insulator-to-metal) of the CF model and found that within the error of numerical calculations both exponents are equal 
to $\nu =1.4\pm 0.2$. We have used optimization procedure in order to determine not only the best fit of the data leading to the critical exponent, but also 
to look for a critical disorder if the energy is fixed. We have found that the value of the critical disorder is energy independent (for energies $\epsilon$ not very close to zero). Obviously, we cannot rule out completely some possible weak dependence of the critical disorder on the energy, especially when $\epsilon\rightarrow\infty$, but for a wide range of energies studied, the critical disorder is $W_{cr}=0.19 \pm 0.02$.

\begin{acknowledgments}
We thank A. L. Chudnovskiy and F. Evers for fruitful discussions.
This research was supported
by the SCE internal research grant, one of us (V.K.) was also partly supported by the BSF grant No. 2006201.
\end{acknowledgments}

\end{document}